\documentclass[runningheads,a4paper]{llncs}

\PassOptionsToPackage{hyphens}{url}
\usepackage{hyperref}
\usepackage{graphicx}
\usepackage{calc}
\usepackage{amssymb}
\usepackage{amstext}
\usepackage{amsmath}
\usepackage{url}
\usepackage{placeins}
\usepackage{multirow}
\usepackage{caption}
\usepackage{subcaption}
\usepackage{hhline}

\usepackage{listings}
\begin{document}

\title{Robustness analysis of Deep Sky Objects detection models on HPC}

\titlerunning{Robustness analysis of Deep Sky Objects detection models}

\author{Olivier Parisot, Diogo Ramalho Fernandes}
\authorrunning{O. Parisot et al.}
\institute{
	Luxembourg Institute of Science and Technology (LIST) \\ 
	5 Avenue des Hauts-Fourneaux, \\
	4362 Esch-sur-Alzette, Luxembourg \\ 
	\email{olivier.parisot@list.lu}
}


\maketitle              
\begin{abstract}
Astronomical surveys and the growing involvement of amateur astronomers are producing more sky images than ever before, and this calls for automated processing methods that are accurate and robust. Detecting Deep Sky Objects -- such as galaxies, nebulae, and star clusters -- remains challenging because of their faint signals and complex backgrounds. Advances in Computer Vision and Deep Learning now make it possible to improve and automate this process. In this paper, we present the training and comparison of different detection models (YOLO, RET-DETR) on smart telescope images, using High-Performance Computing (HPC) to parallelise computations, in particular for robustness testing.
 
\end{abstract}

\section{Introduction}
\label{sec:intro}

Astronomical imaging is expanding rapidly, fueled by major professional surveys and the growing involvement of amateur astronomers \cite{buxner2021amateur}. 
This flood of data makes it essential to develop automated detection methods that stay accurate and robust across diverse conditions, instruments, and noise levels. 
The spread of affordable smart telescopes further increases the value of these tools and results for the amateur astronomy community.

Building on previous works \cite{parisot2024deepspaceyolodataset,parisot2024deep} that demonstrated the potential of Deep Learning architectures for Deep Sky Objects (DSO) detection in images captured from smart telescopes (such as YOLOv7, Pix2Pix) -- DSO are celestial objects located beyond our solar system -- such as galaxies, nebulae, and star clusters -- that appear as extended, diffuse sources through telescopes, distinguishing them from point-like sources like individual stars or solar system objects.

Although the apparent positions of DSO in the sky are well known and can ultimately be identified in images by comparison with existing catalogues \cite{steinicke2010observing}, the main advantage of automatic algorithms lies in enhancing the efficiency of observation systems. 
This can be achieved either by assisting in device calibration (such as camera settings, focusing, etc.) or by guiding the observer -- for example, by indicating unfavorable observing conditions (if a target is not visible when it should be, etc.).
Currently, detection models are also used for educational and even recreational purposes to feed a YouTube channel dedicated to Electronically Assisted Astronomy (EAA) \cite{parisot2025method}.

In this context, the present study extends the analysis to additional recent state-of-the-art models, including recent versions of YOLO and RET-DETR. 
More precisely, testing newer YOLO versions beyond v7 is interesting, as they often bring architectural and training improvements that can enhance detection accuracy and robustness. 
RET-DETR, a Transformer-based model, offers advanced context understanding that may better handle faint or complex astronomical objects \cite{lv2023detrs}.

While a comprehensive comparison of YOLO models for this particular task has recently been carried out on a related dataset \cite{ramos2025deep}, the present study places particular emphasis on evaluating the robustness of the trained models to various noise types, notably by testing their performance on datasets where perturbations are introduced in a controlled and gradual manner.

To support these experiments, High-Performance Computing (HPC) resources were employed to conduct large-scale, parallelised training and evaluation under varied noise conditions.

In Section \ref{sec:related}, we review related work on robusness analysis for Deep Learning detection methods.
We brievly present the data, including acquisition process with smart telescopes in Section \ref{sec:data}.
We describe the training procedures and model architectures in Section \ref{sec:training}.
We evaluate the models’ performance and robustness to various noise types and levels in Section \ref{sec:robustness}.
Finally, we discuss the implications of our findings in Section \ref{sec:discussion}, and we propose directions for future work in Section \ref{sec:conclusion}.

\section{Related works}
\label{sec:related}

To assess and compare the effectiveness of object detection models, standard evaluation metrics such as precision, recall, and mean Average Precision (mAP) are typically employed. 
These metrics provide quantitative insights into a model’s ability to correctly identify and localize objects across predefined test datasets.
But for a more in-depth comparison, in particular to assess the robustness of the models, other approaches can be used, such as adversarial perturbations and common corruptions.

The impact of introducing various types of noise (Gaussian, Poisson, Salt and Pepper, Speckle) in the Military Aircraft Detection dataset to test a YOLOv5 model has been analysed in \cite{bakir2023evaluating}.
Going further with the same idea, the authors of \cite{rodriguez2024impact} have estimated the effect of noise and brightness on object detection methods for a subset of the COCO dataset.
In \cite{apostolidis2025delving}, a research team has checked adversarial attacks on YOLOv3 to YOLOv11 models pretrained on the COCO dataset too.
We can also cite a study that applied a range of common corruptions -- including various noise types, blurs, and more -- to evaluate the robustness of detection methods on aerial imagery \cite{he2024robustness}.

In the field of astronomy, although not directly related to detection, it is worth noting that a recent study investigated the effect of applying adversarial perturbations on a ResNet18-based galaxy morphology classification model using data from the Large Synoptic Survey Telescope (LSST) \cite{ciprijanovic2021robustness}.

In this paper, we focus on training 16 detection models using a dataset of images captured by smart telescopes and annotated with the present DSOs, and we test how the trained models respond when varying levels of noise are added during the test phase.

\section{Data}
\label{sec:data}

As data source, we have employed the DeepSpaceYoloDataset, originally introduced in \cite{parisot2024deepspaceyolodataset}, which contains 4696 RGB astronomical images captured by two smart telescopes and annotated with the precise positions of visible DSO (such as galaxies, nebulae, and star clusters). 
The dataset spans observations from March 2022 to September 2023 across three European locations -- Luxembourg, France, and Belgium -- specifically chosen to represent urban light-pollution environments. 
The images were captured using two smart telescopes (Stellina and Vespera) — the acquisition protocol is detailed in the data description paper \cite{parisot2024deepspaceyolodataset}.

Formatted in the standard YOLO layout, DeepSpaceYoloDataset includes matched JPEG images and annotation text files, making it fully compatible with state-of-the-art object detection pipelines. 
This curated dataset is highly suitable for training and evaluating Deep Learning models tailored to the detection of celestial objects under real-world noise and visibility conditions.

\section{Models training}
\label{sec:training}

In the initial work presented in \cite{parisot2024deepspaceyolodataset,parisot2024deep}, YOLOv7 models as well as other indirect detection techniques based on XRAI \cite{kapishnikov2019xrai} and even Pix2Pix were presented, yielding satisfactory results.

Here, the first step was to retrain a variety of models using the DeepSpaceYoloDataset, via different versions of YOLO (5, 8, 11, 12) \cite{ali2024yolo}, ranging from very small models to very large models, but also testing RET-DETR \cite{lv2023detrs}, which is supposed to be effective.

To this end, data (images and labels) was organized as follows:
\begin{itemize}
\item Training set: files names starting with 1, 2, 3 and 4 -- i.e. 4252 images.
\item Validation set: files names starting with 5, 6 and 7 -- i.e. 333 images.
\item Test set: files names starting with 8 and 9 -- i.e. 222 images.
\end{itemize}

The training process was designed and performed for 400 epochs, by using a Python script based on Ultralytics \footnote{\url{https://www.ultralytics.com/}}.
A total of 16 different models were considered, and to train these models efficiently, each computation was launched with SLURM on GPU-nodes of the MeluXina HPC \footnote{\url{https://docs.lxp.lu/first-steps/handling_jobs/}}.
The HPC compute nodes each have 512GB of RAM and are equipped with 4 A-100 GPUs, which allowed us to take advantage of the multi-GPU capabilities offered by Ultralytics \footnote{\url{https://docs.ultralytics.com/fr/modes/train/\#multi-gpu-training}}. 
This was useful for the larger models (e.g. YOLO12x, 59.1M parameters), and we were also able to benefit from a large batch size during training (32).

Data augmentation was applied during the training, including random flips, scaling, translations, and backgrounds to increase robustness under different sky conditions (default parameters of Ultralytics were used).

Models accuracies were evaluated by using the standard object detection metrics:
\begin{itemize}
    \item \textbf{Precision} = $\frac{TP}{TP + FP}$: proportion of predicted DSO that are correct.
    \item \textbf{Recall} = $\frac{TP}{TP + FN}$: how many actual DSOs were correctly detected.
    \item \textbf{mAP@50} (mean Average Precision at IoU threshold 0.50): detection performance with a relaxed overlap requirement.
    \item \textbf{mAP@50–95}: mAP over multiple IoU thresholds (from 0.50 to 0.95, step of 0.05), reflecting both localization and classification accuracy.
\end{itemize}

\begin{table}[]
\begin{center}
\caption{Accuracy of models on the test set of DeepSpaceYoloDataset.}
\label{table:results}
\begin{tabular}{|l|l|l|l|l|}
\hline
Model         & precision     & recall        & mAP50         & mAP50-95      \\
\hline
retdetr-dso-l & 0.76          & 0.58          & 0.65          & 0.45          \\
retdetr-dso-x & 0.75          & 0.58          & 0.65          & 0.44          \\
\hline
yolo-dso-5n   & 0.80          & 0.64          & 0.73          & 0.53          \\
yolo-dso-5m   & 0.79          & 0.64          & 0.74          & 0.54          \\
yolo-dso-5x   & 0.78          & 0.70          & 0.76          & 0.57          \\
\hline
yolo-dso-8n   & 0.78          & 0.61          & 0.72          & 0.54          \\
yolo-dso-8m   & 0.77          & 0.67          & 0.74          & 0.56          \\
yolo-dso-8x   & \textbf{0.88} & 0.59          & 0.73          & 0.55          \\
\hline
yolo-dso-11n  & 0.78          & 0.64          & 0.74          & 0.57          \\
yolo-dso-11m  & 0.77          & 0.67          & 0.74          & 0.55          \\
yolo-dso-11x  & 0.77          & 0.65          & 0.73          & 0.54          \\
\hline
yolo-dso-12n  & 0.83          & 0.63          & 0.75          & 0.55          \\
yolo-dso-12s  & 0.84          & 0.68          & 0.78          & 0.60          \\
yolo-dso-12m  & 0.79          & \textbf{0.72} & \textbf{0.78} & \textbf{0.61} \\
yolo-dso-12l  & 0.81          & 0.67          & 0.75          & 0.60          \\
yolo-dso-12x  & 0.77          & 0.66          & 0.75          & 0.58          \\
\hline
\end{tabular}
\end{center}
\end{table}

Considering the first set of results (Table~\ref{table:results}), which shows how the detection models perform (with NMS enabled) on the DeepSpaceYoloDataset test set, and training the models in a fairly similar manner (same data, same number of epochs, same batch size), we can see that YOLO8x has the best precision, while YOLO12m achieves the best recall, the best mAP50, and the best mAP50-95.
This ultimately demonstrates that the improvements introduced in YOLO12 are generally very interesting for this specific use case, while also noting that RET-DETR techniques lag significantly behind here.
It is also worth mentioning that YOLOv5-based versions also yield good results.

\section{Robustness analysis}
\label{sec:robustness}

Given the inherently noisy nature of astronomical images, the DeepSpaceYoloDataset test set intrinsically contains a non-negligible level of noise, but we want to check if the trained models are capable of making accurate detections when the noise is higher.
We have therefore re-evaluated the models on different versions of the DeepSpaceYoloDataset test dataset, applying various degrees of noise, and then collected the performance metrics (precision, recall, mAP50, mAP50@95).

There are several types of noise that can be found in images \cite{boncelet2009image}.
Three types of noise were tested during our experiemnts, as they commonly appear in astronomical images:
\begin{itemize}
\item Gaussian noise, caused by camera electronics or thermal effects, adds small random variations to each pixel, making the image look grainy. 
\item Poisson noise, or photon noise, comes from the discrete nature of light, with brighter areas showing more fluctuations than darker ones. 
\item Salt-and-Pepper noise, caused by sensor errors or transmission problems, randomly turns some pixels completely black or white, creating scattered spots.
\end{itemize}

To this end, we wrote a Python script to load test data (images and expected annotations), add noise to images by keeping the same annotations, compute the predictions with the detection models, and then evaluate the metrics.

The following Python code was used to add Gaussian noise, Poisson nosie and Salt and Pepper noise in images:

\begin{lstlisting}
def add_gaussian_noise(img, level):
    noise = np.random.normal(0, level, img.shape).astype(np.float32)
    noisy_img = np.clip(img.astype(np.float32) + noise, 0, 255).astype(np.uint8)
    return noisy_img

def add_poisson_noise(img, level, max_level=25):
    if level <= 0: return img.copy()
    img_norm = img.astype(np.float32) / 255.0
    scale_for_poisson = 50
    noisy = np.random.poisson(img_norm * scale_for_poisson) / scale_for_poisson
    alpha = level / max_level  
    img_out = (1 - alpha) * img_norm + alpha * noisy
    img_out = np.clip(img_out * 255.0, 0, 255).astype(np.uint8)
    return img_out

def add_salt_and_pepper(img, level):
    if level<=0: return img.copy()
    p=np.clip(level/100,0,0.25);
    out=img.copy();
    r=np.random.rand(*img.shape[:2])
    out[r<p/2]=255;
    out[r>1-p/2]=0
    return out
\end{lstlisting}

Given that the number of different combinations was quite high, sequential execution of the Python script with dedicated parameters would have taken a very long time.
As in the model training phase, computations were executed in parallel on GPU-nodes of the MeluXina HPC using SLURM job arrays, with one batch assigned to each noise type and noise level.

\begin{figure}[!h]
        \center{\includegraphics[width=0.85\textwidth]{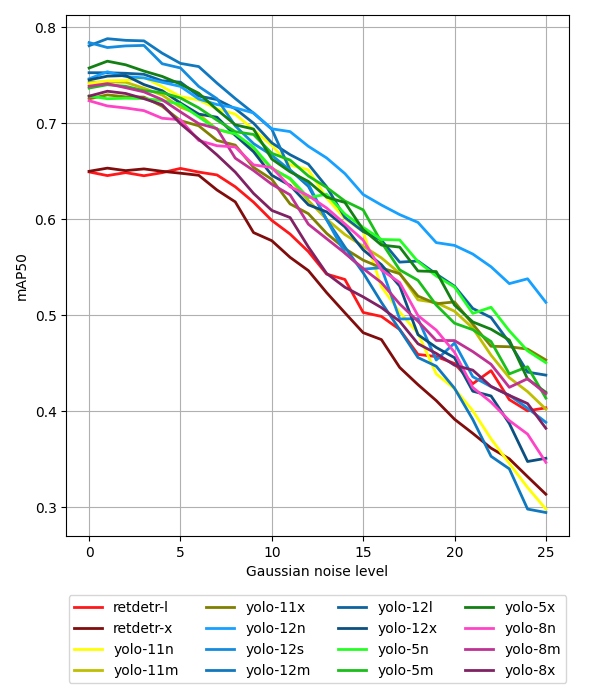}}
        \caption{Effect of Gaussian noise on the detection models accuracies: the x axis show the level of noise applied on the test set of DeepSpaceYoloDataset and the y axis shows the corresponding mAP50. }
        \label{fig:gaussian}
\end{figure}

\begin{figure}[!h]
        \center{\includegraphics[width=0.85\textwidth]{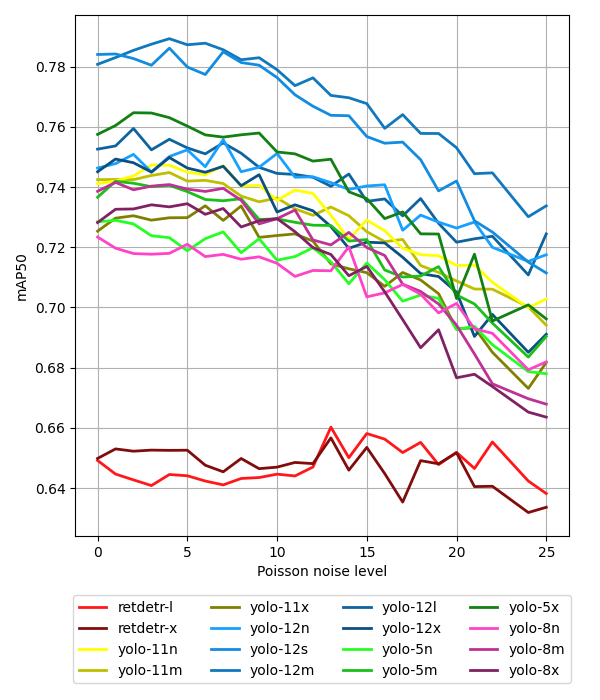}}
        \caption{Effect of Poisson noise on the detection models accuracies: the x axis show the level of noise applied on the test set of DeepSpaceYoloDataset and the y axis shows the corresponding mAP50. }
        \label{fig:poisson}
\end{figure}

\begin{figure}[!h]
        \center{\includegraphics[width=0.85\textwidth]{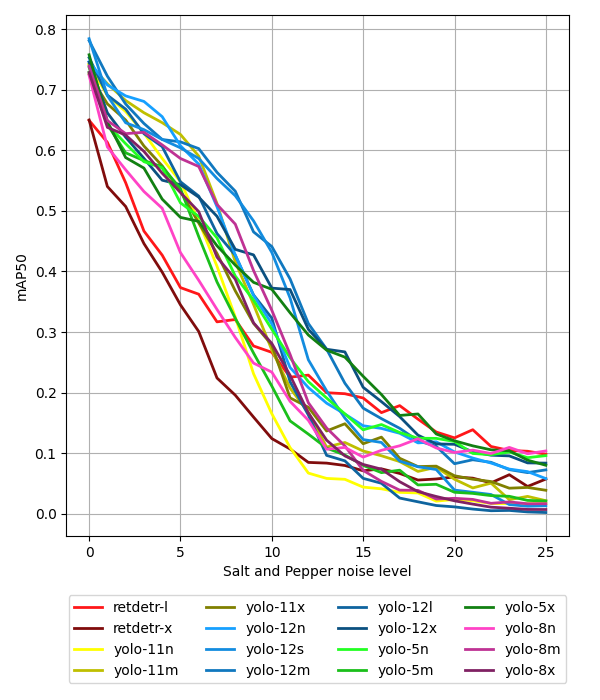}}
        \caption{Effect of Salt and Pepper noise on the detection models accuracies: the x axis show the level of noise applied on the test set of DeepSpaceYoloDataset and the y axis shows the corresponding mAP50. }
        \label{fig:saltpepper}
\end{figure}

Across our experiments (Figures \ref{fig:gaussian}, \ref{fig:poisson}, \ref{fig:saltpepper}), it is evident that gradually introducing noise highlights the superior robustness of the YOLO12 family models, with YOLO12n demonstrating the highest resilience to various noise types. 
By comparison, RET-DETR models consistently underperform across most scenarios, except under extreme \textit{Salt and Pepper} noise conditions.

When looking in detail at the evolution of the metrics, we observe, for example, that the progressive addition of Gaussian or Poisson noise significantly decreases recall, while it can sometimes help improve precision (notably for Gaussian noise levels between 10 and 20, and Poisson noise levels between 5 and 10).

While not directly linked to noise, we also examined the effect of JPEG quality variations on the model’s robustness to compression:
\begin{lstlisting}
def add_jpeg_perturbation(img, level):
    if level <= 0: return img.copy()
    _, enc = cv2.imencode(".jpg", img, [int(cv2.IMWRITE_JPEG_QUALITY), 100-level])
    return cv2.imdecode(enc, 1)
\end{lstlisting}

\begin{figure}[!h]
        \center{\includegraphics[width=0.85\textwidth]{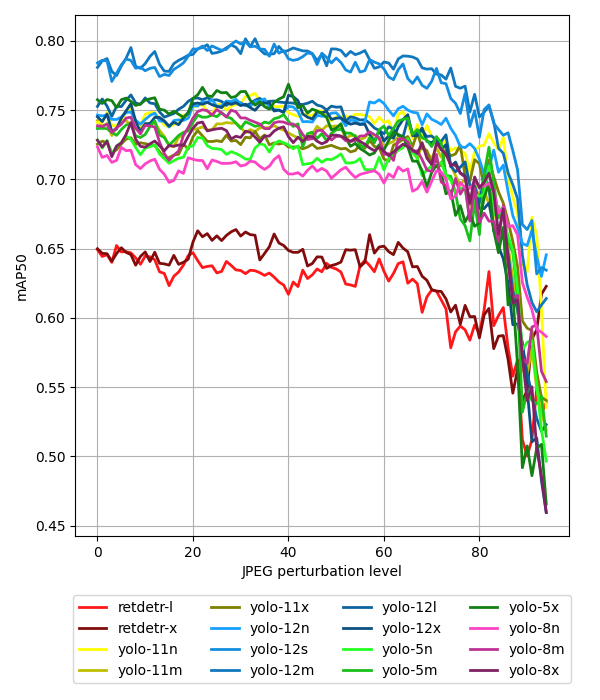}}
        \caption{Effect of JPEG compression artefacts on the detection models accuracies: the x axis show the level of compression degradation applied on the test set of DeepSpaceYoloDataset and the y axis shows the corresponding mAP50. }
        \label{fig:jpeg}
\end{figure}

In Figure \ref{fig:jpeg}, we can see that the models are not particularly sensitive to this type of compression, as long as it is not too severe.
Moreover, the compression does not really change the hierarchy of models in terms of efficiency: it is only possible to see that slight compression may possibly help RET-DETR models to perform better.

\section{Discussion}
\label{sec:discussion}

This paper primarily aims to present the approach we followed to identify the model(s) most suitable for detecting DSOs, while remaining robust enough to adapt to varying situations.
Noise is a particularly significant factor in EAA and astrophotography: although the goal is always to improve the signal-to-noise ratio during data capture, high noise levels can still occur, requiring algorithms that are sufficiently effective to handle them.
In this specific case, the YOLOv12 models we trained proved to be particularly well suited.

Another objective was to lay the groundwork for a toolbox enabling the training and in-depth evaluation of detection models using a HPC infrastructure.
Thanks to the developed scripts (Python and Bash, with SLURM for jobs handling) it is possible to reproduce these tests on new models with the same input/output specifications, as well as to easily add new metrics -- such as robustness tests against background sky variations, walking noise, potential gradients, and so on.
This provides with a solid foundation for continuing the work, which we will likely complement with Explainable AI (XAI) techniques.

We also now have a basis for extending both our datasets and models to other types of telescopes beyond those used to create the DeepSpaceYoloDataset. 
This includes instruments with different focal lengths and apertures, as well as other types of sensors, potentially equipped with interference filters, among others.

\section{Conclusion and perspectives}
\label{sec:conclusion}

In this work, we demonstrated the potential of modern object detection architectures, such as recent YOLO and RET-DETR, for the challenging task of Deep Sky Objects detection in astronomical images captured with smart telescopes. 
Leveraging an HPC infrastructure allowed us to efficiently train and evaluate multiple models under a variety of conditions, providing a basis for performance comparison. 
Our results highlight that YOLO12 models have the best accuracy, and are relatively robust regarding the noise (Gaussian, Poisson, Salt and Pepper), and even JPEG compression.
Conversely, the RET-DETR models appear to be less effective in this case, although the results could probably be improved by tuning the hyperparameters.

In future work, we aim to design and implement dedicated detection models that move beyond standard architectures, while leveraging more extensive parallel computing to further accelerate and enhance the training and evaluation process, all while also focusing on the interpretability of the models’ behavior. \\

\FloatBarrier

\noindent \textbf{Acknowledgements}: This research was funded by the Luxembourg Institute of Science and Technology (LIST), during the NEOD project (\url{https://www.list.lu/en/informatics/project/neod/}). The simulations were performed on the Luxembourg national supercomputer MeluXina. The authors gratefully acknowledge the LuxProvide teams for their expert support. \\ \\

\noindent \textbf{Data availability}: DeepSpaceYoloDataset can be retrieved from this page: \url{https://zenodo.org/doi/10.5281/zenodo.8387070}. Additional materials used to support the results of this paper are available from the corresponding author upon request.

\bibliographystyle{splncs04}
\bibliography{refs}

\begin{thebibliography}{10}
\providecommand{\url}[1]{\texttt{#1}}
\providecommand{\urlprefix}{URL }
\providecommand{\doi}[1]{https://doi.org/#1}

\bibitem{ali2024yolo}
Ali, M.L., Zhang, Z.: {The YOLO framework: A comprehensive review of evolution,
  applications, and benchmarks in object detection}. Computers
  \textbf{13}(12), ~336 (2024)

\bibitem{apostolidis2025delving}
Apostolidis, K.D., Papakostas, G.A.: {Delving into YOLO Object Detection
  Models: Insights into Adversarial Robustness}. Electronics  \textbf{14}(8),
  ~1624 (2025)

\bibitem{bakir2023evaluating}
Bak{\i}r, H., Bak{\i}r, R.: {Evaluating the robustness of YOLO object detection
  algorithm in terms of detecting objects in noisy environment}. Journal of
  Scientific Reports-A (054),  1--25 (2023)

\bibitem{boncelet2009image}
Boncelet, C.: Image noise models. In: The essential guide to image processing,
  pp. 143--167. Elsevier (2009)

\bibitem{buxner2021amateur}
Buxner, S.R., Fitzgerald, M.T., Freed, R.M.: Amateur astronomy: Engaging the
  public in astronomy through exploration, outreach, and research. Space
  Science and Public Engagement pp. 143--168 (2021)

\bibitem{ciprijanovic2021robustness}
{\'C}iprijanovi{\'c}, A., Kafkes, D., Perdue, G., Pedro, K., Snyder, G.,
  S{\'a}nchez, F., Madireddy, S., Wild, S., Nord, B.: Robustness of deep
  learning algorithms in astronomy--galaxy morphology studies. arXiv preprint
  arXiv:2111.00961  (2021)

\bibitem{he2024robustness}
He, H., Ding, J., Xu, B., Xia, G.S.: On the robustness of object detection
  models on aerial images. IEEE Transactions on Geoscience and Remote Sensing
  (2024)

\bibitem{kapishnikov2019xrai}
Kapishnikov, A., Bolukbasi, T., Vi{\'e}gas, F., Terry, M.: {XRAI: Better
  attributions through regions}. In: Proceedings of the IEEE/CVF international
  conference on computer vision. pp. 4948--4957 (2019)

\bibitem{lv2023detrs}
Lv, W., Xu, S., Zhao, Y., Wang, G., Wei, J., Cui, C., Du, Y., Dang, Q., Liu,
  Y.: {DETRs Beat YOLOs on Real-time Object Detection} (2023)

\bibitem{parisot2024deepspaceyolodataset}
Parisot, O.: {DeepSpaceYoloDataset: Annotated Astronomical Images Captured with
  Smart Telescopes}. Data  \textbf{9}(1), ~12 (2024)

\bibitem{parisot2025method}
Parisot, O.: {Method and Tools to Collect, Process, and Publish Raw and
  AI-Enhanced Astronomical Observations on YouTube}. Electronics
  \textbf{14}(13), ~2567 (2025)

\bibitem{parisot2024deep}
Parisot, O., Jaziri, M.: {Deep Sky Objects Detection with Deep Learning for
  Electronically Assisted Astronomy}. Astronomy  \textbf{3}(2),  122--138
  (2024)

\bibitem{ramos2025deep}
Ramos, L.T., Rivas-Echeverr{\'\i}a, F.: {Deep sky object detection in
  astronomical imagery using YOLO models: a comparative assessment}. Neural
  Computing and Applications pp. 1--23 (2025)

\bibitem{rodriguez2024impact}
Rodriguez-Rodriguez, J.A., L{\'o}pez-Rubio, E., {\'A}ngel-Ruiz, J.A.,
  Molina-Cabello, M.A.: The impact of noise and brightness on object detection
  methods. Sensors  \textbf{24}(3), ~821 (2024)

\bibitem{steinicke2010observing}
Steinicke, W.: {Observing and cataloguing nebulae and star clusters: from
  Herschel to Dreyer's New General Catalogue}. Cambridge University Press
  (2010)

\end{thebibliography}

\end{document}